\documentstyle[epsfig]{article}
\begin{document}
\begin{center}
  {\Large{\bf How Crucial is Small World Connectivity for Dynamics?}}
  \\
  \vspace{1cm} Prashant M. Gade$^{1,2}$ and Sudeshna
  Sinha$^{1}$\\$^1$ {\em The
    Institute of
    Mathematical Sciences, Taramani, Chennai 600 113, India}\\
$^2${\em Centre for
    Modelling and Simulation, University of Pune, Ganeshkhind, Pune,
    411 007, India}
\end{center}

\vspace{1cm}

\begin{abstract}
  We study the dynamical behaviour of the collective field of chaotic
  systems on small world lattices. Coupled neuronal systems as well as
  coupled logistic maps are investigated. We observe that significant
  changes in dynamical properties occur only at a reasonably high
  strength of nonlocal coupling. Further, spectral features, such as
  signal-to-noise ratio (SNR), change monotonically with respect to
  the fraction of random rewiring, i.e. there is no optimal value of
  the rewiring fraction for which spectral properties are most
  pronounced. We also observe that for small rewiring, results are
  similar to those obtained by adding small noise.


\end{abstract}


\section{Introduction}

The inadequacy of treating coupled systems as finite dimensional
lattices on one hand and fully random networks on the other, has
become evident in recent times. Various networks, ranging from
collaborations of scientists to metabolic networks, have been studied
and shown not to fit in either paradigm.  Some alternatives have been
suggested, the most popular of which are small-world networks
\cite{Watts} and scale free networks \cite{barabasi}. In the small
world model, one starts with a structure on a lattice, for instance
regular nearest neighbour connections. Then each link from a site to
its nearest neighbor is rewired randomly with probability $p$, {\it
  i.e.}  the site is connected to another randomly chosen lattice
site. This model is proposed to mimic real life situations in which
non-local connections exist along with predominantly local
connections. The geometrical properties of these lattices have been
extensively studied. Many studies have observed the following
\cite{relev}: starting from a one dimensional chain at $p=0$, one
obtains long-range order at any finite rewiring probability with same
critical exponents as in the mean-field case, namely in the
thermodynamic limit the behavior for any $p\neq 0$ is the same as the
behavior for $p=1$ for these models. Newman and Moore recover critical
exponents for percolation on small world lattices which are the same
as for the Bethe lattice, i.e. an infinite dimensional case
\cite{Newman}.  For the XY model, Medvedyeva {\it et al} conjecture
that the critical exponents are the same as for the mean field
case\cite{medve}. They have confirmed it for $p \geq 0.03$ and there is
good reason to believe that it is true for any $p>0$ (The obvious
difficulty is that one needs to simulate larger and larger lattices at
small $p$.) Similar conclusions are reached for the Ising model on
small world networks as well \cite{hong}.

So while there is much evidence that random nonlocal connections, even
in a small fraction, makes a big difference to geometrical properties
like characteristic path length, its implications for dynamics is
still unclear and even conflicting. So the first question we will
probe here is this: {\em does one see dynamical changes at very low
  values of $p$}, namely does the behavior change as soon as non-local
shortcuts are put in place (as observed in equilibrium models).

Now, while the dynamics of coupled oscillators and coupled maps on
regular lattices has been extensively investigated, there have been
only a few studies on the dynamics on small world lattices. Most of
these have focussed on exact synchronization. For instance, in Coupled
Map Lattices (CML), with add-on non-local links, it was observed that
synchronization occurs in the thermodynamic limit for infinitesimal
$p$ \cite{gade}. Likewise Barahona {\it et al} \cite{pecora}
investigated coupled oscillators with small world connections with
add-on links, and observed that in cases where synchronization occurs,
the fraction $p$ of nonlocal edges required to synchronize the system
decreases monotonically with lattice size and is very small for large
lattices. However, for CMLs where non-local links were added at the
cost of existing regular links, transition to exact synchronization
was observed at finite $p$ \cite{sinha, Jost}. This finite $p$
transition was similar to the transition to self-sustained
oscillations evident in a model of infection spreading
\cite{abramson}. It was also shown in these CMLs that the magnitude of
lyapunov exponents and the coupling parameter range over which exact
synchronization occurs, varies monotonically with $p$ \cite{sinha}.
Further, a study on coupled Chate-Manneville minimal maps \cite{cos}
indicates that the critical exponents for the transition to turbulence
change monotonically as a function of rewiring probability $p$. Here
one can find situations in which the critical exponent drops to zero
at some value of $p$, thus changing the nature of the transition from
second order to first order, though the change is still monotonic.
Another investigation on stochastic resonance on small world networks
falls on similar lines \cite{Gao}.  So all these studies indicate that
dynamical features vary monotonically with $p$, and interpolate
between the limits of regular and random connections without in any
sense being ``optimal'' or more pronounced at some intermediate value
of $p$ \cite{Campos}.

Surprisingly however, a few studies indicate a non-monotonic change in
dynamics and special features at small values of $p$. He {\it et al}
reported that the spirals which were unstable on regular networks get
stabilized at very small $p$ \cite{hugang}.  However, it is known that
small spatial noise stabilizes spirals and small world connectivity
could be playing the same role\cite{xin} since the nonlocal
connections will join sites well inside one phase with the other phase
and vice versa. Another interesting study by Lago-Fernandez {\it et
  al} demonstrated that the power spectra of the collective field of
coupled Hodgin-Huxley elements shows a non-monotonic increase in
strength of spectral peaks at low frequency. Thus they argued that
small world connectivity gives something special which is amiss in
regular or random lattices \cite{Lago}. Though this trend held true
for their particular model system, it is not clear if small world
connectivities will have similar consequences in general. Numerical
evidence from more varied sources is required, in the absence of
analytical results, to settle this question.

In view of the above, the second interesting question we wish to
address through our case studies here is as follows: is there evidence
for dynamical features at some intermediate value of $p$, that, in
some sense, does not interpolate between the random and regular
limits. Thus this paper will attempt to provide some more examples
from coupled dynamical systems, including another prototypical
neuronal model, in order to {\em shed more light on the validity of
  the conjecture that small world connections yield special dynamical
  features that are absent in both the regular and the random limits}.

Note that the change in characteristic length scales under small world
connectivities would make a significant change in the characteristic
time scales in geometric models, such as those describing epidemics or
rumor propagation. In these models the initial disturbance is
localized and the time taken for spreading is reduced considerably in
presence of nonlocal connections. In coupled chaotic systems,
however, the characteristic time scale will be related to largest
lyapunov exponent which does not change drastically with nonlocal
connections. Also note that in these spatiotemporally chaotic systems,
the disturbances are neither localized nor few. This is another
important distinction from epidemic models.

Our test systems are the following two networks: (i) coupled
Hindmarsh-Rose neurons and (ii) coupled logistic maps. We will study
both systems in the parameter regime that shows chaos. Note that the
constituents of the networks in the two case studies are very
different. One of them is an excitable system, while other becomes
chaotic via familiar period-doubling route to chaos.

In particular, we will focus on the spectral properties of the
collective field under varying degrees of random rewiring in the two
networks. The mean field indicates the degree of independence of
different elements in the network. If the elements are uncorrelated
and individually chaotic, one would expect the mean field to approach
a constant, namely the average value of the components, with the
fluctuations decaying with system size. On the other hand if the
elements are very coherent, we may see strong departures from this
behaviour. Thus we may look at signal-to-noise ratio (SNR) of the
peaks in the spectrum of the collective field as some kind of {\em
  order parameter} demonstrating the coherent oscillations in the mean
field, if any. It is zero when there are no coherent oscillations in
the mean field while it has a finite value when it oscillates with
some chosen frequency.

As mentioned above, we will explore two questions in this work. First,
we will examine if dilute rewirings have any significant impact on the
spectral properties. Secondly we will try to discern whether or not
any non-monotonic changes result in dynamical properties as rewiring
fraction $p$ is varied, namely we will address the question: does
there exist an optimal $p$ for which certain dynamical features become
significantly more pronounced than in either the regular or random
limits.

The organization of the paper is as follows. In Section 2 we report on
our first test case, namely a network of coupled Hindmarsh-Rose
neurons. In section 3, we report our results for a network of coupled
logistic maps in the chaotic regime. We summarize our results in
Section 4.

\section{Network of Coupled Hindmarsh-Rose neurons}

In light of the observation that the low frequency spectral peak of
the collective field of a network of Hodgkin-Huxley neurons is more
pronounced for low $p$ than for either the regular or the random case
\cite{Lago}, we have chosen our first case study to be another
prototypical neuronal model: the Hindmarsh-Rose model. On lines of
Lago-Fernandez {\it et al} \cite{Lago}, let us cosider a lattice made
up of non-identical Hindmarsh-Rose neurons \cite{hind}. Let each of
them be in chaotic regime and be coupled to its neighbours This system
is defined as

\begin{eqnarray}
{\frac{dx_{ij}}{dt}}&=&y_{ij} +3 x_{ij}^2 -x_{ij}^3-z_{ij} +I_{ij}-\epsilon
\Delta
\\
{\frac{dy_{ij}}{dt}}&=&1-5x_{ij}^2-y_{ij}
\\
{\frac{dz_{ij}}{dt}}&=&-rz_{ij}+rS(x_{ij}+1.6)
\end{eqnarray}
where $\Delta$ is a proportional to laplacian of $x$ variable
calculated at site $(i,j)$. For a regular square lattice,
$\Delta=4*x_{ij}-x_{i1,j}-x_{i2,j}-x_{i,j1}-x_{i,j2}$ where $j1=j-1$,
$j2=j+1$, $i1=i+1$, $i2=i-1$ (with periodic boundary conditions).
However for a small world lattice $i1,i2,j1,j2$ could take random
values with probability $p$.
 
The parameters in the above equations take values: $r=0.0021,
\epsilon=0.1$ and $S=4$. The term $I_{ij}=3.281\pm 0.025 \eta$ where
the $\eta$ is a random number between 0 and 1.

The collective (mean) field in the above system of coupled neurons can
be defined as

$$h = \frac{1}{N} \sum_{i,j} x_{ij}$$

We study this quantity for different system sizes $N$ and different
rewiring fractions $p$.

\begin{figure}[htb]
\label{fig0}
\begin{center}
\mbox{\epsfig{file=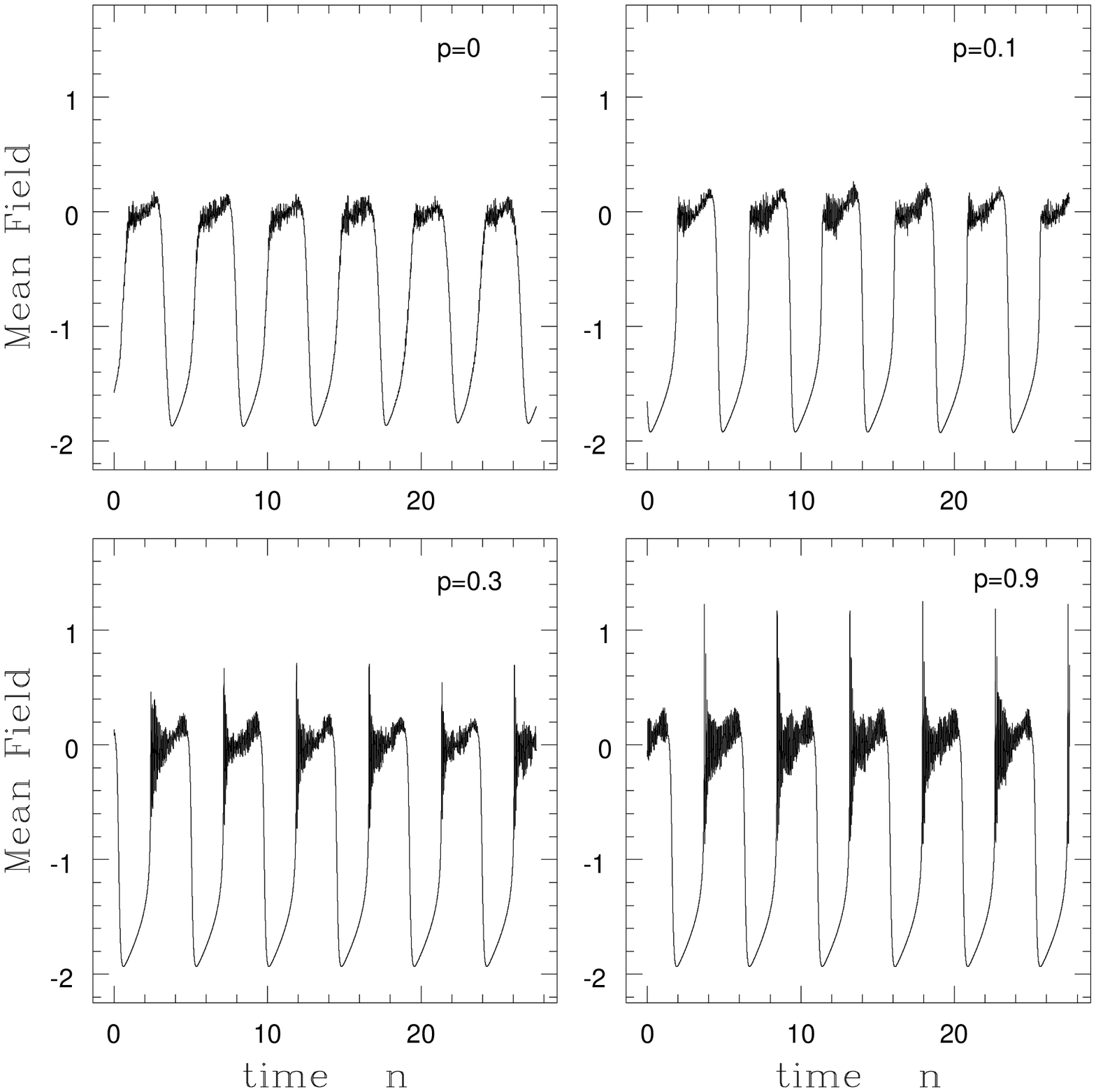,width=15truecm}}
\end{center}
\caption{Time evolution of the collective field of networks of coupled
  Hindmarsh-Rose neurons, after a transience of $10^4$, with rewiring
  fraction $p =$ (a) 0 (b) 0.1 (c) 0.3 and (d) 0.9. The system size is
  $20 \times 20$.}
\end{figure}

Fig.~1 displays the time evolution of the collective field for four
values of rewiring fraction $p$. It seems apparent that there is a
monotonic change in the qualitative nature of the dynamics. The trend
is as follows: as $p$ increases, i.e. as the network becomes more and
more random, the oscillations of the mean field get larger in
amplitude and follow the pattern of the individual neurons more
closely. This indicates increasing synchronicity as $p \to 1$, though
exact synchronisation is never achieved here.

The spectra of the collective field reveal the following trends:

(a) For a regular lattice ($p \sim 0$) the peak in the power spectrum
of the collective field occurs only in the low frequency region.

(b) For fully nonlocal connections ($p \to 1$) the power
spectrum of the mean field shows peaks both in the low as well as the
high frequency regimes.

(c) As one changes the rewiring probability $p$ we see more and more
peaks in high frequency regime. But the {\em low frequency peak is
  neither destroyed nor decreased in strength}. In fact, it increases
faster than background noise giving a slightly higher signal to noise
ratio.

(d) The background level of the power spectrum grows as $p$ is
increased. This is keeping in with expectation, since for higher $p$
the sites will be more correlated.

In Fig. 2 we illustrate the above points by showing the power spectra
for the representative cases of $p=0$, $p=0.1$ and $p=0.999$.

\begin{figure}[htb]
\label{fig1}
\begin{center}
\mbox{\epsfig{file=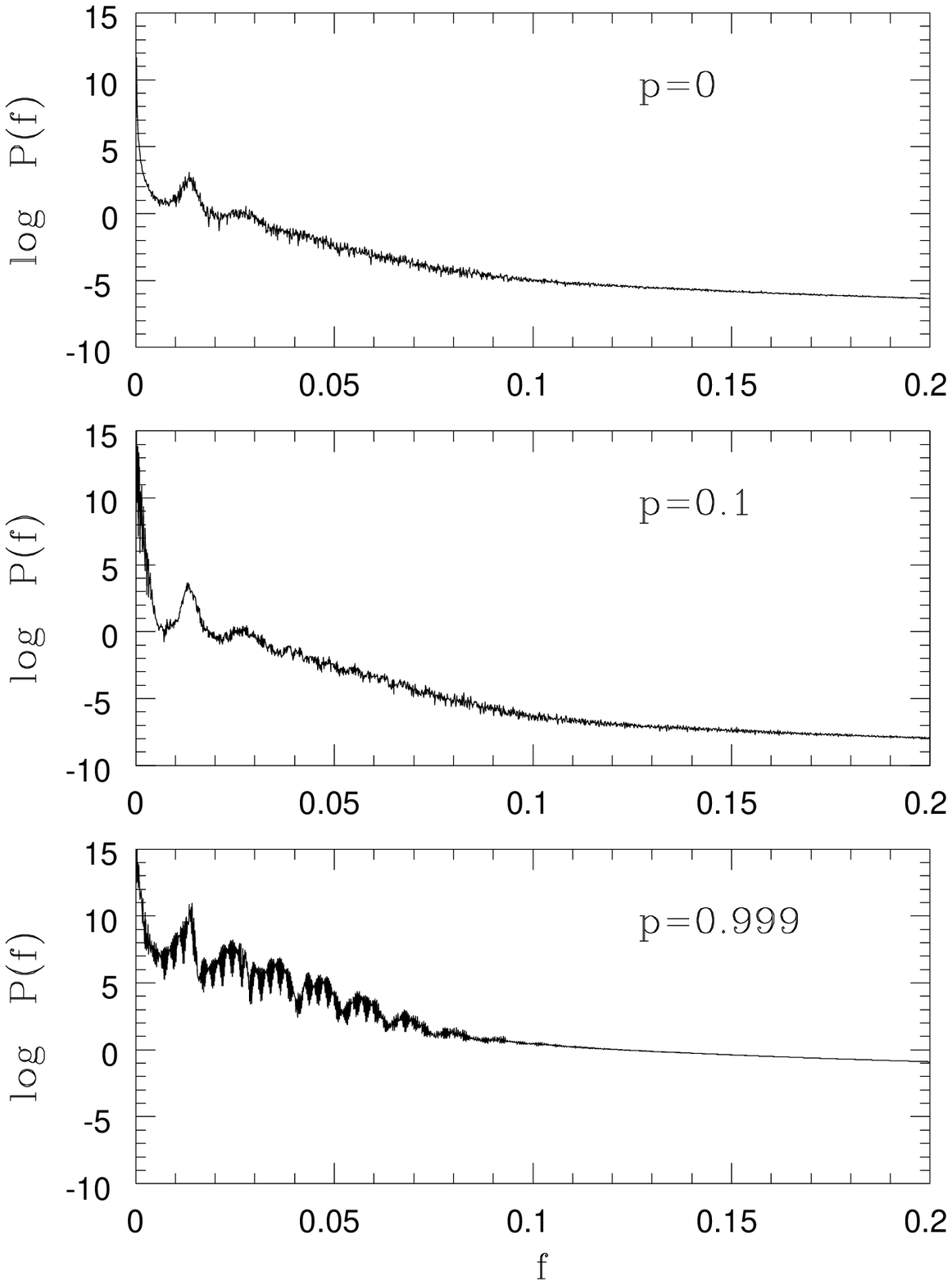,width=15truecm}}
\end{center}
\caption{ Power spectra of the collective field of networks of coupled
  Hindmarsh-Rose neurons, with rewiring fraction $p =$ (a) 0 (b) 0.1
  and (c) 0.999. Here we average over $25$ time series runs of $8192$
  data points each. The time step is $0.01$ in the fourth order
  Runge-Kutta algorithm employed to evolve the system, and the field
  at every tenth step is sampled. Thus the longest time scale picked
  up is $\sim 10^3$. The system size is $151 \times 151$. The
  frequencies are scaled by $0.1$.}
\end{figure}


In light of the remarks made in introduction, we would like to compare
the system with rewired nonlocal couplings with a system with regular
couplings influenced by noise. So we simulate the above dynamics on a
regular lattice under the influence of noise, i.e. we evolve the
dynamics above with an additive noise term $\sigma \eta$, where $\eta$ is a
random number between -1 and 1.

The spectra under different noise strengths $\sigma$ shows the
following trends:

(a) Upto noise strengths $\sigma < 0.001$ the spectral peak remains the
same, and arguably even increases a little. So the effect of very
small noise is akin to low rewiring fractions ($p \to 0$).

(b) For larger noise strengths $\sigma > 0.001$, the spectral peak
decreases significantly. So here higher noise strengths do not have
the same effect as higher rewiring fractions ($p \to 1$).

(c) As noise increases, the $P(f) \sim 1/f^2$ background gets cleaner
and more pronounced.

Fig.~3 illustrates these trends in a lattice of size $N=151\times 151$.

\begin{figure}[htb]
\label{fig2}
\begin{center}
\mbox{\epsfig{file=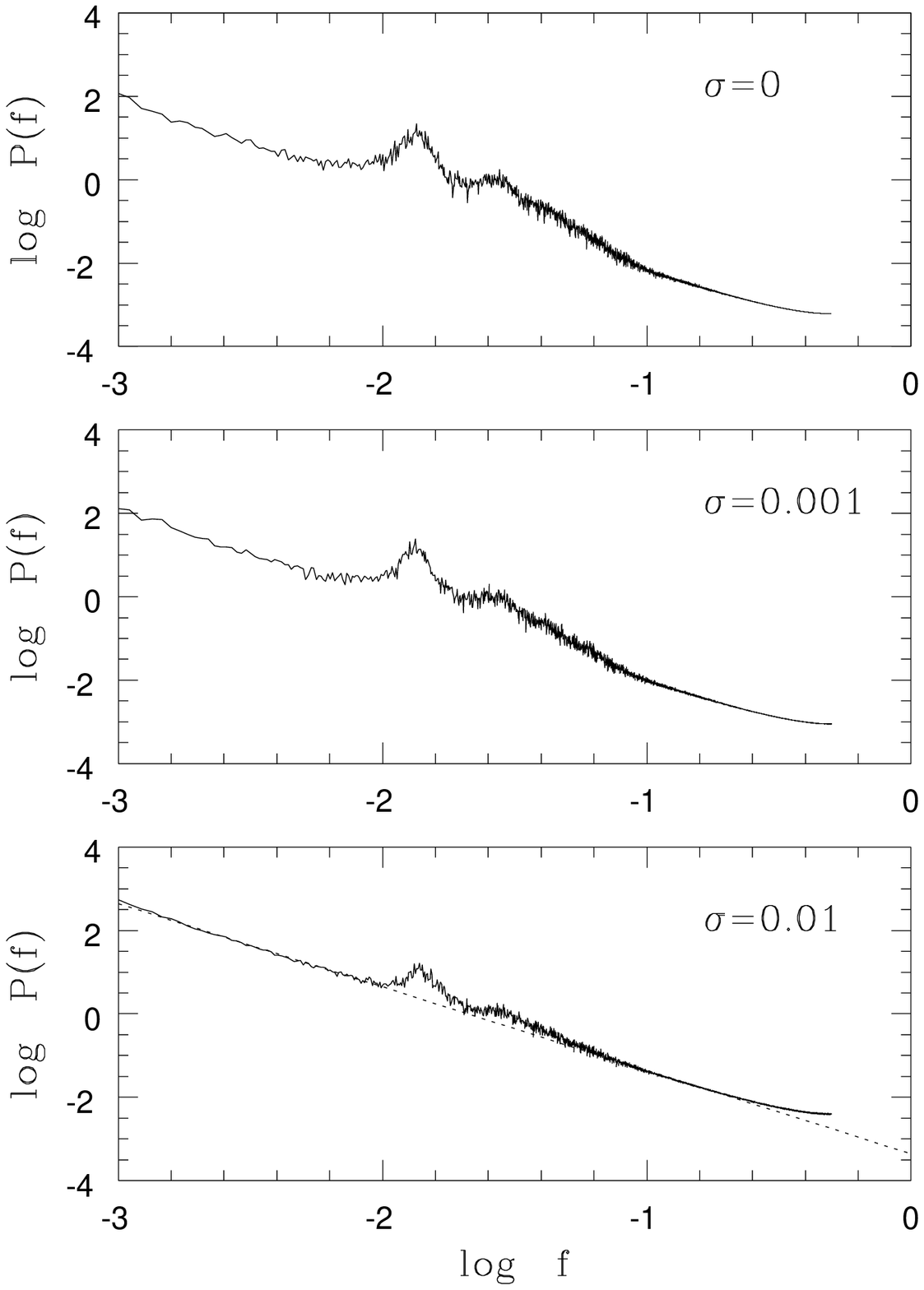,width=15truecm}}
\end{center}
\caption{Power spectra of the collective field of regular ($p=0$) 
  networks of coupled Hindmarsh-Rose neurons, with noise strengths
  $\sigma =$ (a) 0 (b) 0.001 and (c) 0.01 (with the dotted $P(f) = 1/f^2$
  line also displayed). Here we average over $25$ time series
  runs of $8192$ data points each. The time step is $0.01$ in the
  fourth order Runge-Kutta algorithm employed to evolve the system,
  and the field at every tenth step is sampled. The system size is
  $151 \times 151$.  The frequencies are scaled by $0.1$.}
\end{figure}

We have also studied the system with the connections dynamically
rewired, i.e. at very small intervals the connectivity matrix is
updated keeping the probability of rewiring fixed at $p$. Dynamic
rewiring yields results qualitatively similar to static rewiring for
small $p$. At larger $p$ however, dynamic rewiring is unlike static
rewiring. In fact it's effects are rather similar to that of noise at
larger $\sigma$. So in dynamically rewired networks, as $p$ increases, the
rough oscillatory behaviour of the collective field is rapidly lost
and the spectra is dominated by the $1/f^2$ background (see Fig.~4).

\begin{figure}[htb]
\label{fig1}
\begin{center}
\mbox{\epsfig{file=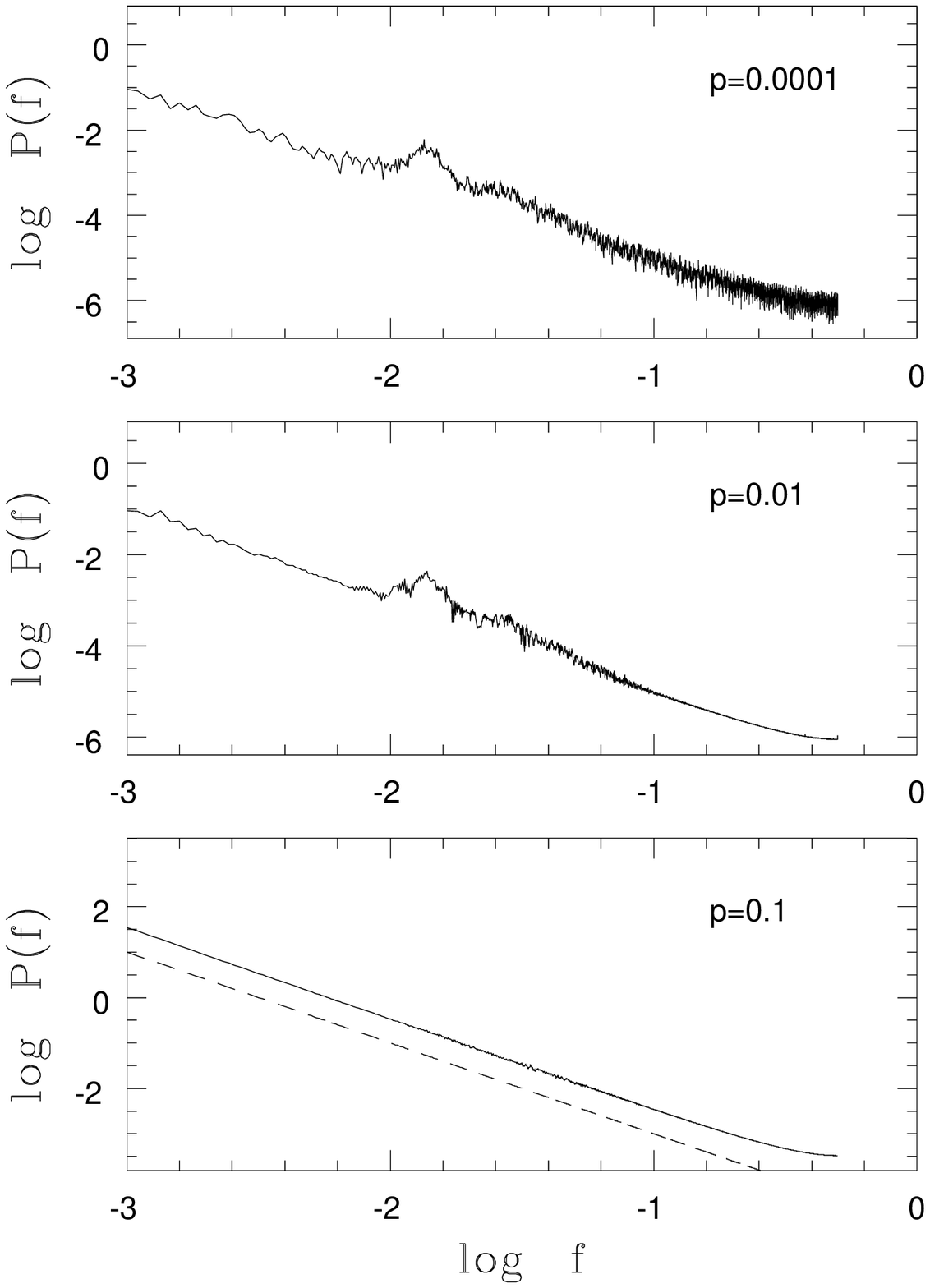,width=15truecm}}
\end{center}
\caption{ Power spectra of the collective field of  
  coupled Hindmarsh-Rose neurons, with connections dynamically rewired.
  Three rewiring fractions are displayed : (a) $p = 0.0001$ (b) $p =
  0.01$ and (c)$p = 0.1$ (with the dashed $P(f) = 1/f^2$ line also
  shown).  Here we average over $25$ time series runs of $8192$ data
  points each. The time step is $0.01$ in the fourth order Runge-Kutta
  algorithm employed to evolve the system, and the field at every
  tenth step is sampled. The system size is $151 \times 151$.  The
  frequencies are scaled by $0.1$.}
\end{figure}

Various other questions could be asked about dynamics on networks,
like for instance the nature of dynamic phase transitions. Such
questions would involve concerns about the relevant order
parameter(s), thermodynamic limit in space and asymptotic limit in
time and delicate issues in approaching these limits\cite{grass}. But
here we will not concern ourselves with the asymptotic or
thermodynamic limit, and we will draw no conclusions about infinite
networks from our studies of finite (albeit quite large) lattices.
Indeed from a practical point of view, whether it is a collection of
neurons or electronic oscillators, our observations will still be very
relevant.\\

\section{Network of coupled logistic maps}

We have also studied coupled chaotic logistic maps on a 2-dimensional
small world network. This system is given as

\begin{eqnarray}
x_{n+1} (i,j) = (1-\epsilon) f(x_n (i,j)) +
 \frac{\epsilon}{4} \{ f(x_n (i1,j)) + f(x_n (i2,j)) \\ \nonumber
\ \ \ \ \ \ \ \ \ \ \ \ \ \ \ \ \ \ \ \ + f(x_n(i,j1)+f(x_n(i,j2)\}
\end{eqnarray}
Again $i1,i2,j1,j2$ takes the nearest neighbour values: $i+1$, $i-1$,
$j+1$ and $j-1$ with probability $1-p$, and could take random values
with probability $p$.  The network has periodic boundary conditions.
We carried out simulations for function $f(x)=1-rx^2$ for $r=1.82$,
$r=1.9$ and $r=2$.

We obtained the mean field, again defined as $\frac{1}{N} \sum_{i,j}
x(i,j)$, for the above system, with $N=201\times 201$. We have the
following observations, for both static and dynamic rewiring:

\begin{figure}[htb]
\label{fig3}
\begin{center}
\mbox{\epsfig{file=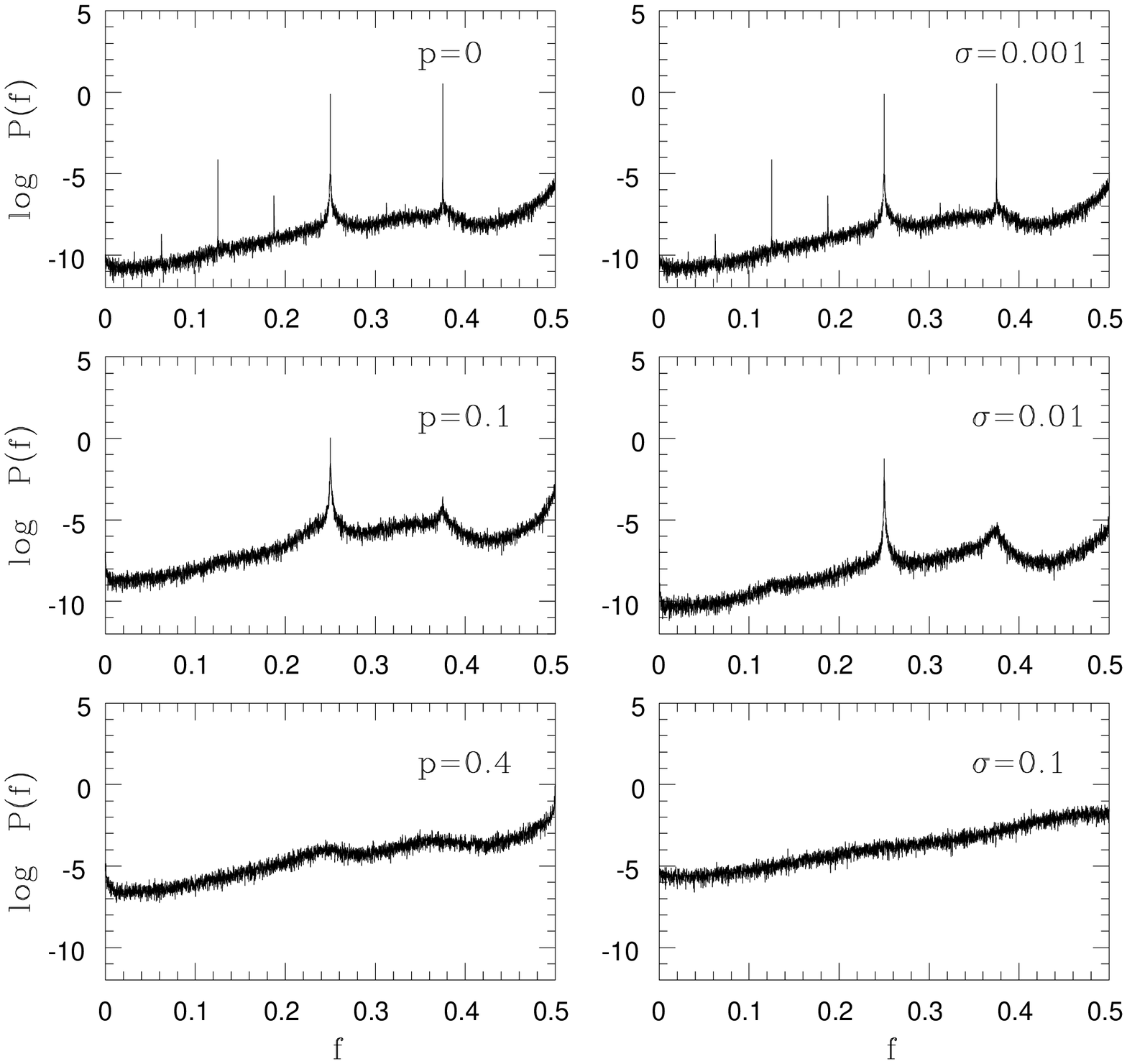,width=15truecm}}
\end{center}
\caption{ Power spectra of the collective field of  
  coupled logistic map networks: (left) with rewiring fraction $p =$
  (a) 0 (b) 0.1 and (c) 0.4; and (right) with $p=0$ and noise
  strengths $\sigma =$ (a) 0 (b) 0.01 and (c) 0.1 in Eq.~5. Here system
  size $N = 350 \times 350$ and we average over $25$ time series runs of
  $8192$ data points each.}
\end{figure}

(a) For $r=2$, the spectum does not have sharp peaks and spectral
features do not change much with rewiring of bonds.

(b) For $r=1.82$ and $r=1.9$, there are $\delta$ peaks for regular
lattices, and as one increases the rewiring probability $p$ the
spectral peaks reduce in strength and ultimately vanish.

(c) The background level grows as $p$ is increased, as expected.

(d) The SNR displays monotonic decrease with respect to rewiring
probability $p$.

In the left panel of Fig.~5 we display the power spectrum of the above
system with $r=1.82$ in the local maps, for $p=0$, $p=0.1$ and
$p=0.4$. The figure clearly bears out the observations listed above.
If the spectral peaks are not very pronounced to begin with at $p=0$,
even very small rewiring probability $p$ will destroy it. However, if
the spectral peaks are sharp at $p=0$, one has to go up to larger
values of $p$ to see them vanish.

In Fig.~6 we display the signal-to-noise ratio (SNR) of the spectra at
different values of rewiring fraction $p$, for the case of $r=1.82$.
The figure illustrates that the SNR is a monotonic function of $p$. So
evidently no significant dynamical effect is discernable as $p \to 0$.
Rather the SNR shows a gradual decrease with increasing $p$.

\begin{figure}[htb]
\label{fig1}
\begin{center}
\mbox{\epsfig{file=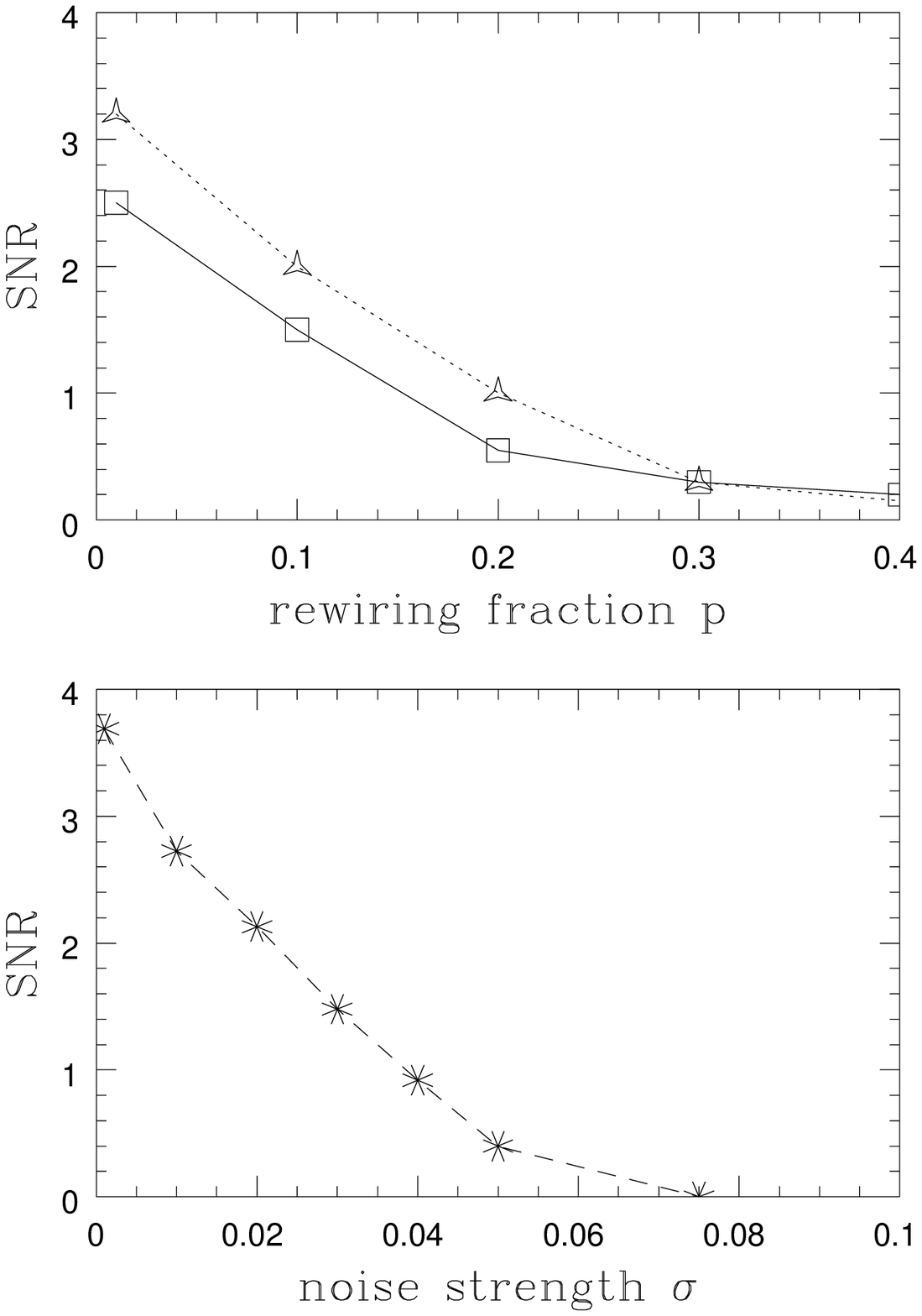,width=15truecm}}
\end{center}
\caption{The upper panel displays the signal-to-noise ratio (SNR) 
  of the power spectra of the collective field of coupled logistic maps
  for different values of rewiring fraction $p$. The lower panel
  displays the SNR of the power spectra for a regular network under
  different noise strengths $\sigma$.  The system size is $20 \times 20$.}
\end{figure}

We have also simulated the dynamics of a coupled logistic map network
on a regular lattice under varying noise strengths, namely:
\begin{eqnarray}
x_{n+1} (i,j) = (1-\epsilon) f(x_n (i,j)) +
 \frac{\epsilon}{4} \{ f(x_n (i+1,j)) + f(x_n (i-1,j)) \\ \nonumber
\ \ \ \ \ \ \ \ \ \ \ \ \ \ \ \ \ \ \ \ + f(x_n(i,j+1)+f(x_n(i,j-1) \}
+ \sigma \xi
\end{eqnarray}
where $\xi$ is a random number in the range $[-1:1]$ and $\sigma$ is
the noise strength, and interestingly we observed a very similar
diminishing of spectral peaks with increasing $\sigma$.

In the right panel of Fig.~5 we display the power spectrum of the
collective field in the above system with $r=1.82$ in the local maps,
for noise strengths $\sigma=0.001$, $\sigma=0.01$ and $\sigma=0.1$.
Clearly the noise destroys the spectral peaks much in the same way as
increasing $p$, as evident in the close similarity between the left
and right panels of Fig.~5. This indicates that non-local connections
act as spatial noise in this dynamical network, with rewiring fraction
$p$ playing the role of noise strength $\sigma$.

We have also studied the system with the connections dynamically
rewired, i.e. at every iteration $n$ the connectivity matrix is
updated keeping the probability of rewiring fixed at $p$. Dynamic
rewiring yields results qualitatively similar to static rewiring for
small $p$. However in a dynamically rewired network the SNR falls much
more sharply than for static rewiring. For instance Fig.~7 displays
the spectra for dynamic rewiring for the case of $r=1.82$, $N = 150 \times
150$. Clearly the peaks have vanished for $p=0.1$ when the connections
are dynamically rewired, while they disappear only around $p \sim 0.4$
when the rewired connections are static.

\begin{figure}[htb]
\label{fig1}
\begin{center}
\mbox{\epsfig{file=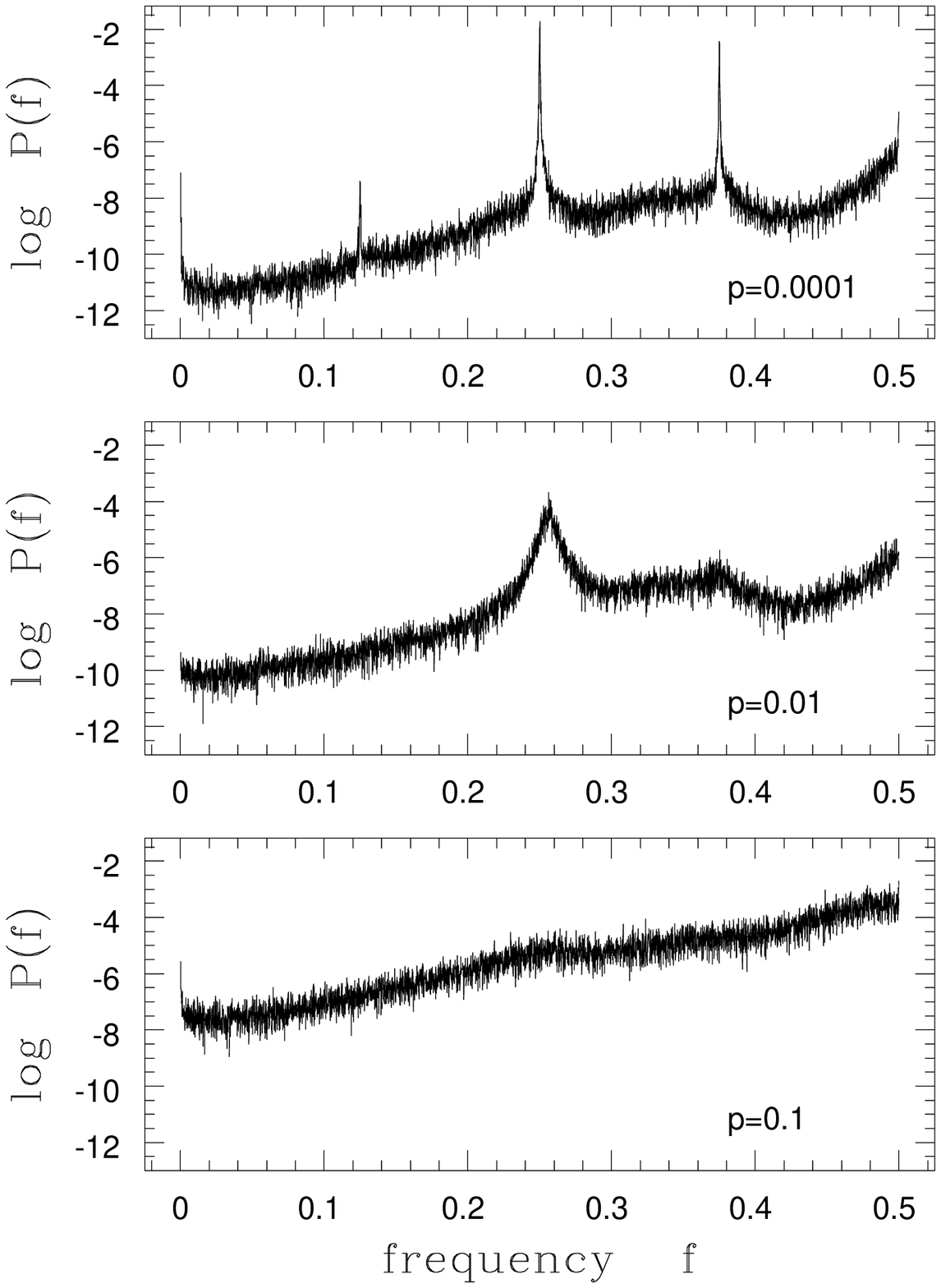,width=14truecm}}
\end{center}
\caption{ Power spectra of the collective field of  
  coupled logistic map networks with connections dynamically rewired.
  Three rewiring fractions $p =$ (a) 0.0001 (b) 0.01 and (c) 0.1 are
  displayed. Here system size $N = 350 \times 350$ and we average over $25$
  time series runs of $8192$ data points each.}
\end{figure}

These results arising from a very different system further strengthens
our conclusion that $p \to 0$ does not have special implications for
dynamical properties. Rather the dynamical characteristics appear to
change smoothly and monotonically with respect to rewiring fraction
$p$.

\section{Conclusions} 

It had been observed in a study of the collective field of coupled
Hodgin-Huxley neurons that in the small world region the low frequency
spectral peak was more pronounced than it was in either the fully
regular or fully random case \cite{Lago}. Our first objective here was
to check this feature in another prototypical neuronal model in order
to gauge the generality and range of applicability of the above
phenomena. So we chose a system of coupled Hindmarsh-Rose neurons as
our first case study.

The key results of the study on coupled Hindmarsh-Rose neurons showed
that the change in the low frequency spectral peak as a function of
random rewiring $p$ is {\em monotonic}. There was no evidence of any
significant increase in spectral strength in the low $p$ regime. So
the Hindmarsh-Rose neuron network, unlike the Hodgin-Huxley neuron
network of Ref. \cite{Lago}, does not yield special dynamical features
in the dilute limit of small world links. In fact the nature of the
dynamics of the mean field appears to vary quite smoothly and
monotonically throughout the full range of $p$.

In our second case study we studied a network of coupled logistic
maps. Here too the key result was that the spectral features changed
monotonically with respect to $p$ and did not show any prominent
change at any special value of intermediate $p$. We also found that
the dynamics at small rewiring was very similar to that of regular
connections with additive noise. 

These observations then provide examples in support of the conjecture
that in a large number of coupled dynamical systems the changes in
collective behaviour is monotonic with respect to the degree of
non-locality in connections, and is not in any way ``optimised'' at any
particular $p$.

\end{document}